\documentclass[conference,10pt]{IEEEtran}
\IEEEoverridecommandlockouts

\usepackage{cite}
\usepackage{amsmath}
\usepackage{tikz}
\usetikzlibrary{tikzmark, positioning}
\usepackage{tabularx}
\usepackage{graphicx}

\usepackage[colorlinks]{hyperref}
\usepackage{algorithmic}

\usepackage{float}

\usepackage{array}

\usepackage{graphicx}

\usepackage{svg}

\ifCLASSOPTIONcompsoc
 \usepackage[caption=false,font=footnotesize,labelfont=sf,textfont=sf]{subfig}
\else
 \usepackage[caption=false,font=footnotesize]{subfig}
\fi
\usepackage{fixltx2e}

\usepackage{stfloats}
\usepackage{url}

\def\BibTeX{{\rm B\kern-.05em{\sc i\kern-.025em b}\kern-.08em
    T\kern-.1667em\lower.7ex\hbox{E}\kern-.125emX}}

\begin{document}

\makeatletter

\def\ps@IEEEtitlepagestyle{%
  \def\@oddhead{\mycopyrightnotice}%
  \def\@evenhead{}%
  \def\@oddfoot{}%
  \def\@evenfoot{}%
}

\def\mycopyrightnotice{%
  \begin{minipage}{\textwidth}
  \centering
  \footnotesize
  © 2024 IEEE. Personal use of this material is permitted. Permission from IEEE must be obtained 
  for all other uses, in any current or future media, including reprinting/republishing 
  this material for advertising or promotional purposes, creating new collective works, 
  for resale or redistribution to servers or lists, or reuse of any copyrighted component 
  of this work in other works.
  \end{minipage}%
  \gdef\mycopyrightnotice{}
}

    \newcommand{\linebreakand}{%
      \end{@IEEEauthorhalign}
      \hfill\mbox{}\par
      \mbox{}\hfill\begin{@IEEEauthorhalign}
    }


\makeatother
%
\title{HPAC-IDS: A Hierarchical Packet Attention Convolution for Intrusion Detection System}




\author{
    \vspace{3mm}
    \IEEEauthorblockN{\textbf{Anass Grini}\IEEEauthorrefmark{1},
                      \textbf{Btissam El Khamlichi}\IEEEauthorrefmark{1},
                      \textbf{Abdellatif El Afia}\IEEEauthorrefmark{2},
                      \textbf{Amal El Fallah-Seghrouchni}\IEEEauthorrefmark{1}\IEEEauthorrefmark{4}}
    \vspace{3mm}
    
    \IEEEauthorblockA{\IEEEauthorrefmark{1}Ai movement, The International Artificial Intelligence Center of Morocco, UM6P, Rabat, Morocco}
    \IEEEauthorblockA{\IEEEauthorrefmark{4}Lip6, Sorbonne University, Paris, France}
    \IEEEauthorblockA{\IEEEauthorrefmark{2}ENSIAS, Mohammed V University, Rabat, Morocco}

    \thanks{For the published version, see: \href{https://doi.org/10.1109/WCNC57260.2024.10570804}{WCNC57260.2024.10570804}}
}



%


\setlength{\extrarowheight}{1pt}

\maketitle
\vspace*{-1.05cm}
\begin{abstract}

This research introduces a robust detection system against malicious network traffic, leveraging hierarchical structures and self-attention mechanisms. The proposed system includes a \textit{Packet Segmenter} that divides a given raw network packet into fixed-size segments that are fed to the HPAC-IDS. The experiments performed on CIC-IDS2017 dataset show that the system exhibits high accuracy and low false positive rates while demonstrating resilience against diverse adversarial methods like Fast Gradient Sign Method (FGSM), Projected Gradient Descent (PGD), and Wasserstein GAN (WGAN). The model's ability to withstand adversarial perturbations is attributed to the fusion of hierarchical attention mechanisms and convolutional neural networks, resulting in a 0\% to 10\% adversarial attack severity under tested adversarial attacks with different segment sizes, surpassing the state-of-the-art model in detection performance and adversarial attack robustness.

\end{abstract}


%
\IEEEpeerreviewmaketitle

\section{Introduction}


In network security, intrusion detection is vital in protecting digital systems and user privacy. However, enhancing these systems is challenging due to hackers' ever-changing tactics to bypass security measures. This ongoing tug-of-war between cybersecurity professionals and adversaries not only drives improvements in security systems but also refines the techniques used by attackers. IDSs, traditionally rely on signature databases to compare incoming network packets with previously reported threats based on their signatures \cite{otoum2021ids}. However, this approach encounters limitations when faced with unknown threats that have not yet been documented in the database. 
Intelligent IDSs were proposed as an improvement to the traditional ones. Intelligent IDSs use Machine Learning (ML) and Deep Learning (DL) techniques to detect malicious traffic efficiently.


ML algorithms learn and recognize intricate patterns in data using K-Nearest Neighbors (KNN), Support Vector Machine (SVM), and Decision Tree \cite{tsai2009intrusion}, enhancing IDS’s ability to accurately distinguish between normal and malicious activities \cite{kim2022robust}. Also, they can adapt to the evolving threat landscape by updating and retraining models \cite{chiche2021towards}.
For DL-based IDSs, different approaches were explored. Some researchers viewed network intrusion as a time-related event, suggesting a time-series approach based on Attention-LSTM neural network \cite{yang2020method, tan2019neural}. Meanwhile, others employed Convolutional Neural Networks (CNNs) to treat network traffic as images for malware classification \cite{wang2017malware, halbouni2022cnn}.

A different approach started to emerge, viewing network traffic as text. Packet2Vec \cite{goodman2020packet2vec} employs a shallow neural network and the word embedding (Word2Vec) methodology to generate packet vectors from n-grams.
As an improvement, \cite{hassan2021intrusion} proposed \textit{PayloadEmbeddings}, utilizing byte embeddings of network payloads. Combined with a shallow neural network and KNN for classification, this system achieved high accuracies across various datasets, surpassing other techniques. The superiority of \textit{PayloadEmbeddings} can be attributed to its substantial vocabulary size and longer vector length, allowing it to capture a more extensive range of contextual information embedded within bytes. However, the authors stated that \textit{PayloadEmbedding} needs to be retrained when there's a change in the nature of attack traffic over time, which makes it vulnerable to new attack types and adversarial samples.

Building on prior advancements in intrusion detection, we propose a packet-embedding approach that considers network packets in their raw form, transcending traditional feature extraction boundaries. This design not only seeks to achieve high accuracy with fewer false positives but also aims to be resilient against adversarial samples intended to bypass detection. In summary, the following contributions are made in this work:

\begin{itemize}
    \item[-] Introduction of a Hierarchical Packet Attention Convolution System (HPAC) tailored for network packets, which innovatively views packets akin to natural language, paving the way for advanced packet analysis.
    
    \item[-] A \textit{packet segmenter} that processes raw packets into fixed-size segments, transforming intricate packet data into an analyzable format, likened to 'sentences' and 'words'.
    
    \item[-] We demonstrated the performance of the HPAC-IDS over State-of-The-Art DL-based IDS with the dataset in \cite{sharafaldin2018intrusion}, highlighting its advanced malicious traffic detection through the integration of hierarchical attention with convolutional networks.
    \item[-] Proven robustness of the HPAC-IDS against advanced adversarial attacks like PGD, FGSM and Wasserstein GAN, signifying its robustness and potential as a frontline defense in the ever-evolving network security landscape.
\end{itemize}


The rest of the paper is structured as follows: Section \ref{section:1} delves into the contemporary state-of-the-art in ML and DL-based IDSs. Section \ref{section:2} introduces the Packet segmenter and the HPAC model. Section \ref{section:3} details the dataset used, the evaluation approach, and setup specifics and presents our experimental findings. The discussion, conclusion, and potential avenues for future research are encapsulated in Section \ref{section:5}.

\section{Related Work} \label{section:1}
Previous works for malicious network packet classification have involved many different techniques in this area. In this section, we focus on methods that apply machine learning and deep learning, and how we can profit from the advancement of natural language classification models to detect malicious network packets.

\vspace{-0.2cm}
\subsection{Machine Learning based IDS} 

Machine Learning Network Intrusion Detection Systems (ML-NIDS) are categorized mainly into \textit{packet-based} and \textit{session-based} types. \textit{Packet-based} ML-NIDS analyzes individual network packets to detect threats, offering high accuracy and low false positives \cite{kim2022real}. However, they might miss attacks masked within standard packets. In contrast, \textit{session-based} ML-NIDS uses statistical data from sessions, efficiently handling large traffic volumes due to consistent feature size, regardless of the session length \cite{kim2022robust}. Yet, these approaches fall short in real-time detection and are better suited for non-immediate threat responses \cite{kim2022real}. The accuracy of ML-NIDS can be influenced by feature selection, but removing critical features can be detrimental \cite{bovenzi2020hierarchical}. A limitation of ML-NIDS is their reliance on specific datasets, potentially affecting their response to new, untrained attack patterns.

\vspace{-0.2cm}
\subsection{Deep Learning based IDS}
Besides these methodologies, researchers tried to leverage the performance of Deep Learning (DL) architectures to handle malicious traffic detection. Previous studies in network packet classification have proposed various techniques. Hwang et al. \cite{hwang2019lstm} introduced an LSTM-based method for packet-level IDS classification, efficiently distinguishing malicious traffic and reducing processing time. While the recurrent network architecture captures sequential network traffic data, it can face issues like vanishing/exploding gradients and inefficient sequential computations. \cite{tan2019neural} addressed these challenges by adding attention mechanisms and positional encoding, enhancing sequence understanding. Their experiments with various attention techniques led to superior attack detection, surpassing Bi-LSTM models. 


Deep learning approaches provide automatic feature extraction without manual feature engineering. However, they require a large amount of data and time to build an effective model against network attacks, and models must take fixed-size input payloads, which leads to a loss in contextual and semantic information \cite{min2018tr}. Hassan et al. \cite{hassan2021intrusion} proposed a payload embedding approach, trained using a shallow neural network, to generate byte embeddings that lead to computing payload vectors used for classification. This payload embedding is fed to a KNN model to evaluate the performance of the embedding model compared to previous ones. One of the shortcomings of this approach is that payload embedding focuses solely on packet payloads for anomaly detection and does not include packet headers, making it vulnerable to header-based attacks.

Li et al. \cite{li2018byte} proposed a Byte Segment Neural Network (BSNN) based on the Hierarchical Attention Network (HAN) architecture. The authors used a hierarchical structure to mirror the document structure on two levels: word-level and sentence-level, assigning different importance to each sentence and word when constructing document representation. They structured raw network datagrams so that each one was transformed into several fixed-length segments that served as inputs to the BSNN. In the same context, Xiao \cite{xiao2021ebsnn} introduced a novel neural network called EBSNN, an Extended Byte Segment Neural Network that classifies the corresponding application or website by examining the first few packets. They suggested a raw packet transformation where different parts were split and processed into fixed-length byte segments. Analogous to text representation in NLP, each byte in the segment is viewed as a character, a segment as a sentence, and the packet can be likened to a document.


\vspace{-0.2cm}
\subsection{From Natural Language to Network packets}
Gao et al. proposed the Hierarchical Convolutional Attention Network (HCAN) model \cite{gao2018hierarchical}, which employed a hierarchical structure and used self-attention mechanisms instead of Recurrent Neural Networks (RNNs) to increase training speed compared to the HAN model, without sacrificing accuracy. The self-attention mechanism was combined with convolutional filters to create a text document classification model that surpassed the current state-of-the-art classifiers while being twice as fast to train.



This paper explores the combined potential of hierarchical structures and self-attention mechanisms, analogizing raw network packets to natural language structures through a \textit{Packet Segmenter}. This approach culminates in introducing a robust malicious network traffic detection system, termed as the \textit{Hierarchical Packet Attention Convolution System} (HPAC-IDS).

\section{Proposed Method}  \label{section:2}

This section briefly presents the architecture of the Hierarchical Packet Attention Convolution System (HPAC) as applied to network packets. The proposed system processes each raw network packet through a \textit{packet segmenter}, producing fixed-size segments comprising 1-byte hexadecimal values. In this context, packet segments are analogized as sentences, while the individual 1-byte hexadecimal within each segment are treated as words. Following this segmentation, the HPAC embeds each packet into a comprehensive representation vector that encapsulates the entirety of the packet's information.

Ultimately, the embedded packet representation is fed into a shallow neural network classifier equipped with a softmax function, which calculates the predicted class of the network packets. This proposed approach to network packet classification demonstrates the potential for leveraging advanced deep learning techniques to effectively process and classify complex data in real-world applications.


\vspace{-0.2cm}

\begin{figure*}[t!]
    \centering
    \includegraphics[scale=0.28]{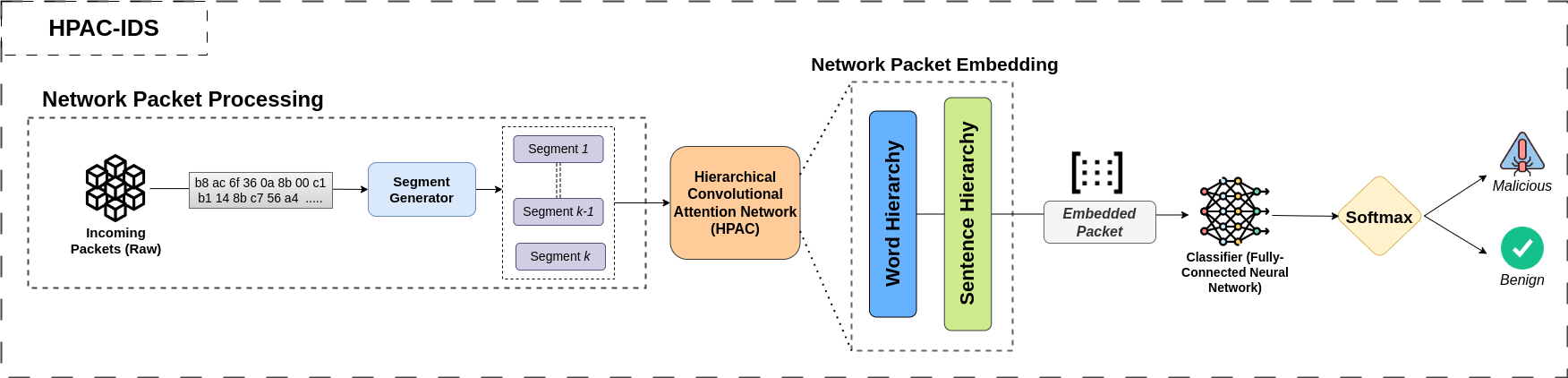}
    \vspace{-0.3cm}
    \caption{General Structure of the HPAC architecture for Malicious Network Packet Detection}
    \label{fig:gen_arch}
    \vspace{-0.5cm}

\end{figure*}

\subsection{Packet Segmenter}

The proposed system views incoming packets as raw, unprocessed hexadecimal strings. The packet segmenter (Figure \ref{fig:seg_generator}) is given these strings, each one is divided by the packet segmenter into smaller, fixed-size segments, each of which contains $k$ single-byte hexadecimal data, ensuring a consistent and manageable data structure for further processing.

Mathematically, the segmented packets can be represented as a sequence of segments $S_i$ as follows:
$$P = S_1,\ S_2,\ ...,S_m,$$
Where $P$ is the incoming packet of length $n$ in terms of hexadecimal values (or bytes), $m$ is the number of segments an incoming packet $P$ will be divided into, which is $m=\lceil \frac{n}{k}\rceil$. The remaining segment $S_m$, which contains $n-k(m-1)$ bytes, is padded with $<PAD>$ values to satisfy equal length for all segments. 

At this stage, we begin to conceptualize each 1-byte hexadecimal value as a distinct word. This approach allows us to draw an analogy between this segmented data and the structure of natural language sentences. In this context, each fixed-size section of $k$-byte hexadecimal values can be considered a sentence made up of $k$ distinct 'words', which are the 1-byte hexadecimal values themselves. By handling the data in this way, we can analyze and manipulate the segmented hexadecimal data more successfully using natural language processing techniques and methodologies.


Each 1-byte hexadecimal value (or 'word') is encoded into a numerical value ranging from 0 to 255 to further process the data. The special padding value, $<PAD>$, is encoded as 256. This encoding transforms the segmented hexadecimal data into a format that's more amenable to computational processing, especially when using machine learning or data analysis techniques.

\begin{figure}[H]
    \centering
    \includegraphics[scale=0.34]{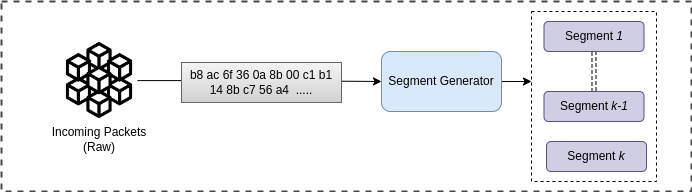}
    \vspace{-0.3cm}
    \caption{Network Packet Segmenter Unit}
    \label{fig:seg_generator}
    \vspace{-0.3cm}
\end{figure}

\subsection{HPAC}
The HPAC architecture, based on HCAN \cite{gao2018hierarchical}, is composed of two primary components: the \textit{Word Hierarchy} and the \textit{Sentence Hierarchy} (as in Figure~\ref{fig:gen_arch}). These hierarchies work in tandem to process and embed the segmented packet data for further analysis and classification. Initially, each word (1-byte hexadecimal) in the segments undergoes an embedding process, after which it is passed to the word hierarchy of the HPAC. Within this hierarchy (Figure \ref{fig:hierarchy}), the word embeddings serve as inputs for convolutional and attention mechanism-based layers, which generate a sentence-level vector representation capturing the relationships and patterns among the words.

    

Subsequently, the same structural approach is applied to the sentences (packet segments), as the sentence embeddings are introduced to the sentence hierarchy of the model. This process involves additional convolutional and attention mechanism-based layers, which function to consolidate the embedded sentences into a comprehensive vector representation of the incoming packet, referred to as the \textit{Packet Embedding}. By leveraging word and sentence hierarchies, the HPAC architecture ensures that the packet embedding effectively captures the intricacies and dependencies within the packet data. The generated packet embedding is then fed into a \textit{softmax} activation function in a shallow neural network. This combination facilitates the prediction of the packet's class, categorizing it as either \textit{malicious} or \textit{benign}.

\begin{figure*}[t!]
    \centering
    \includegraphics[scale=0.41]{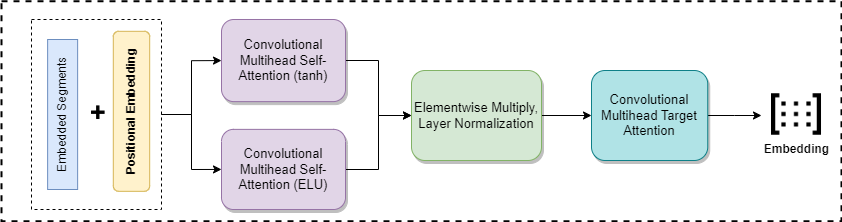}
    
    \vspace{-0.2cm}
    \caption{Hierarchy structure in HPAC}
    \label{fig:hierarchy}
\end{figure*}

\section{Experimental Results} \label{section:3}
\subsection{Dataset}
We used the CIC-IDS2017 \cite{sharafaldin2018intrusion}, a publicly available network intrusion dataset proposed by the Canadian Institute of Cybersecurity (CIC). We start by processing PCAP files using \textit{Scapy}, a Python-based network tool that allows us to parse the different header and payload information from the PCAP files. The publicly available dataset essentially contains five (5) PCAP files, as shown in \ref{table:cicids}. Labeling has been done on the four PCAP files containing malicious traffic. For experiments, the dataset was split in a 60-20-20\%, with 60\% of the instances used for training and 20-20\% for validation and testing, respectively.

\begin{table}[!h]
\centering
\begin{tabular}{|l|p{2.3in}|}
\hline
 \textbf{Day Activity} & \textbf{Attacks Found}  \\
\hline
\footnotesize Monday & \footnotesize Benign (Normal human activities) \\
\hline
\footnotesize  Tuesday & \footnotesize Benign, FTP-Patator, SSH-Patator \\
\hline
\footnotesize  Wednesday & \footnotesize Benign, DoS GoldenEye, DoS Hulk, DoS Slowhttptest, DoS slowloris, Heartbleed \\
\hline
\footnotesize Thursday & \footnotesize Benign, Web Attack – Brute Force, Web Attack – Sql Injection, Web Attack – XSS, Infiltration \\
\hline
\footnotesize Friday & \footnotesize Benign, Bot, PortScan, DDoS \\
\hline
\end{tabular}
\vspace{0.2cm}
\caption{\footnotesize Activity Attacks Included in Each File}
\label{table:cicids}
\vspace{-.6cm}
\end{table}

\vspace{-.2cm}
\subsection{Evaluation Method}
Five standard measures are typically used for the assessment of balanced binary classification: Accuracy (ACC), Detection Rate (DR), Precision, F1 Score (F1), and false-positive rate (FPR). These metrics are defined as follows.

\begin{align*}
        & \fontsize{10pt}{12pt}\selectfont \text{ACC  =  } \frac{TP + TN}{TP + TN + FP + FN}\\[1mm]
        & \fontsize{10pt}{12pt}\selectfont \text{DR  =  } \frac{TP}{TP + FN}\\[1mm]
        & \fontsize{10pt}{12pt}\selectfont \text{FPR  =  } \frac{FP}{TP + TN}\\[1mm]
        & \fontsize{10pt}{12pt}\selectfont \text{Precision  =  } \frac{TP}{TP + FP}\\[1mm]
        & \fontsize{10pt}{12pt}\selectfont \text{F1  =  } \frac{2 \times Precision \times Recall}{Precision + Recall}
\end{align*}


TP (True Positives) represents correctly classified malicious instances, while TN (True Negatives) stands for correctly classified benign instances. FP (False Positives) are instances incorrectly classified as malicious, and FN (False Negatives) are instances incorrectly classified as benign. DR (Detection Rate) reflects our model's accuracy in predicting malicious instances. FPR indicates the false alarm rate when malicious instances are misclassified as benign, and Precision is the ratio of correctly predicted malicious packets to all detected malicious instances.


\subsection{Setup Details}

We conducted all model training and evaluations for our experiments on a high-performance computing system equipped with an Intel(R) Xeon(R) Gold 6152 CPU operating at 2.10GHz. The system utilizes two powerful GPUs, namely the Nvidia(R) Tesla Pascal 40, featuring 22GB of memory, and the Nvidia(R) Tesla VOLTA 100, with 16GB of memory. This hardware configuration enabled us to train and test our models efficiently, ensuring the robustness and reliability of the results we obtained.

Regarding model configuration, we adopted several training hyperparameters to train the HPAC-IDS and EBSNN models to ensure optimal performance. The hyperparameters and their respective values used in the training process are summarized in Table \ref{table:hyperparam}.

We adopted an approach to preprocessing the incoming packet data for efficient network processing and maintaining data integrity. The primary method involved extracting fixed-size segments from each packet. The size of these segments was determined based on the optimal performance of the model, ensuring that the segments were not too small to risk convolution layer errors due to kernel size constraints. Specifically, segment sizes below 6 were avoided as our convolution layers use a kernel size of 3, which could lead to errors for smaller segments.

The models utilize embedding vectors to represent the segments. These embeddings have a dimensionality of 96, which is optimal for capturing semantic relationships between segments without significantly increasing the model's complexity.

The use of Focal Loss is justified by the nature of training data given to the model, it is designed to handle extreme imbalance between classes \cite{lin2017focal}. In our case, Benign packets outnumber malicious ones, which influences the performance of the model during training.


For the HPAC-IDS model, we employ an attention mechanism with 8 heads to capture different aspects and dependencies within the input sequence, allowing the model to recognize complex patterns and relationships.

All models are trained using the Adam optimizer with a learning rate of \(10^{-3}\). The batch size is set to 40, ensuring efficient training while balancing computational resources.

The models are trained over 40 epochs, each consisting of 150 steps. For each step, batches of training data are processed to update the model's parameters. The performance of the models is then evaluated on a validation set, and the best model parameters are saved for final evaluation.

\begin{table}[!h]
\centering
\begin{tabularx}{\linewidth}{|X|c|}
\hline
\footnotesize \textbf{Hyperparameter's name} & \footnotesize \textbf{Values} \\
\hline
\footnotesize Epochs & \footnotesize 40\\
\hline
\footnotesize Loss function & \footnotesize Focal Loss \\
\hline
\footnotesize Optimizer & \footnotesize Adam \\
\hline
\footnotesize Learning Rate & \footnotesize \(10^{-3}\) \\
\hline
\footnotesize Segment Size & \footnotesize 20 \\
\hline
\footnotesize Batch Size & \footnotesize 40 \\
\hline
\footnotesize Embedding Size & \footnotesize 96 \\
\hline
\footnotesize Number of Heads (\textit{for HPAC-IDS model}) & \footnotesize 8 \\
\hline

\end{tabularx}
\caption{\footnotesize HPAC-IDS training hyperparameters}
\label{table:hyperparam}
\vspace{-0.6cm}
\end{table}

\subsection{Experimental Results}

In the course of our study, two models were primarily compared in this study: the \textit{Extended Byte Segment Neural Network (EBSNN}) \cite{xiao2021ebsnn} and our proposed model, \textit{Hierarchical Packet Attention Convolution System (HPAC-IDS)}. The performance of these models was evaluated based on their validation and test metrics.

\begin{table}[!h]
\centering
\small
\begin{tabularx}{\linewidth}{|X|c|c|}
\hline
\textbf{Metrics}  & \textbf{EBSNN} & \textbf{HPAC-IDS} \\ \hline
\footnotesize Validation Accuracy & \footnotesize 0.99500 & \footnotesize \textbf{1.00} \\ \hline
\footnotesize Validation DR & \footnotesize 0.90624  & \footnotesize \textbf{0.99999} \\ \hline 
\footnotesize Validation F1 Score & \footnotesize 0.94959 & \footnotesize \textbf{0.99999} \\ \hline
\footnotesize Validation FPR & \footnotesize 0.08333 & \footnotesize \textbf{0.00} \\ \hline
\footnotesize Validation Loss & \footnotesize 0.61511 & \footnotesize \textbf{3 $\times$ 1e-6} \\ \hline
\footnotesize Validation Precision & \footnotesize 0.90412 & \footnotesize \textbf{0.99999}\\ \hline
\footnotesize Test Accuracy & \footnotesize 0.99905  & \footnotesize \textbf{0.99970} \\ \hline
\footnotesize Test DR & \footnotesize 0.99937  &  \footnotesize \textbf{0.99987}  \\ \hline
\footnotesize Test FPR (\%) & \footnotesize 0.22499 & \footnotesize \textbf{0.02499} \\
\hline
\end{tabularx}
\caption{\footnotesize Performance comparison between EBSNN and HPAC-IDS with segment size $32$}
\label{table:perf}
\vspace{-0.6cm}
\end{table}
\subsubsection{The model precision}
In evaluating the performance metrics between both models on segment size 32, we observe that HPAC-IDS outperforms EBSNN in terms of validation accuracy, detection rate, and F1 score, as depicted in Table \ref{table:perf}. The higher F1 score in particular, indicates that HPAC-IDS is not only accurate but also balanced in terms of precision. This is crucial for IDS, where false negatives and positives can have significant implications.
The notably lower FPR for HPAC-IDS during validation suggests that it is less likely to mistakenly classify benign activities as malicious when compared to EBSNN. This can reduce unnecessary alerts and investigations, making the system more efficient for users.
Furthermore, the test metrics underscore HPAC-IDS's robustness. Its near-perfect accuracy on the test set reaffirms its generalization capabilities. The low values in the FPR on both models during testing affirm the reliability of HPAC-IDS in real-world scenarios.

\begin{table}[!h]
    \centering
    \begin{tabularx}{\linewidth}{|X|c|c|c|c|}
        \hline
        \small \textbf{Work} & \small \textbf{Model} & \small \textbf{Acc \%} & \small \textbf{DR \%} & \small \textbf{FPR \%}\\
        \hline
        \footnotesize Sun et al. \cite{sun2020dl} & \footnotesize CNN+LSTM & \footnotesize 98.67 & \footnotesize 97.21 & \footnotesize 0.47 \\
        \hline
        \footnotesize  Azzaoui et al. \cite{azzaoui2022developing} & \footnotesize  DNN &\footnotesize 99.43  &  \footnotesize 80.33  &  \footnotesize \textbf{0.0007} \\
        \hline
        \footnotesize  Yin et al \cite{yin2023improving} &  \footnotesize Birch + MLP & \footnotesize 99.73    &   -   & \footnotesize 0.15 \\
        \hline
        \footnotesize \textbf{Proposed Method} & \footnotesize \textbf{HPAC} & \footnotesize \textbf{99.79} & \footnotesize \textbf{99.99} & \footnotesize 0.02 \\
        \hline
    \end{tabularx}
    \caption{Comparaison of similar work}
    \label{tab:comp}
\end{table}

In Table \ref{tab:comp}, we compared the performance of our model with other studies that utilized the same dataset. Our analysis reveals that our model outperforms others in terms of both detection rate and accuracy. While our false positive rate (FPR) is not always the lowest, it still demonstrates a relatively strong performance.


\begin{table}[!h]
\centering
\begin{tabularx}{\linewidth}{|X|X|X|}
\hline
\small \textbf{Segment size} & \small \textbf{Test Accuracy} & \small \textbf{Test FPR (\%)} \\
\hline
\footnotesize \textbf{8} & \footnotesize 0.99965  & \footnotesize 0.1499\%\\
\hline
\footnotesize \textbf{20} & \footnotesize 0.99895 & \footnotesize 0.4999\% \\
\hline
\footnotesize \textbf{32} & \footnotesize 0.9997 & \footnotesize 0.0249\% \\
\hline
\footnotesize \textbf{39} & \footnotesize 0.9674 & \footnotesize 16.299\% \\
\hline
\end{tabularx}
\caption{\footnotesize Test Results on different \textit{Segment size}}
\label{table:segSize}
\vspace{-0.6cm}
\end{table}

\subsubsection{The effect of the segment size}

In Table \ref{table:segSize}, we illustrate the performance of the proposed model across varying segment sizes, evaluated based on test accuracy and the false positive rate (FPR). Observing the trend, as the segment size increases from 8 to 32, there's an enhancement in test accuracy, reaching its peak at 0.9997 for a segment size of 32. However, a subsequent increase to a segment size of 39 significantly drops accuracy to 0.9674. Regarding FPR, it starts at 0.1499\% for segment size 8, peaks at 0.4999\% for segment size 20, then sharply drops to 0.0249\% at size 32, only to drastically rise to 16.299\% at size 39. It suggests that while smaller segment sizes might effectively capture crucial features leading to a lower FPR, larger segment sizes, especially beyond 32, might introduce noise or lose essential data patterns, causing a spike in FPR. This data underscores the delicate balance between segment size (granularity) and the model's performance.

\section{Adversarial Robustness} \label{section:4}

To assess the adversarial robustness of the HPAC-IDS and EBSNN models against adversarial attacks, we applied three popular attacks: Projected Gradient Descent (PGD), Fast Gradient Sign Method (FGSM) and Wasserstein GAN (WGAN). Our aim was to measure the deviation of the generated adversarial examples from the original data. We used cosine similarity as a metric to capture this deviation, offering insight into the magnitude of the perturbation and the models' vulnerability to these attacks.

\textit{Adversarial attack severity} was employed as a key metric to assess the models' robustness. This metric reflects the impact of an attack on the models' accuracy and overall performance. The severity of an adversarial attack illustrates the degree to which a model is influenced or misled. It's important to note that severity and accuracy are inversely related: as one rises, the other falls. Leveraging this severity metric is instrumental in developing strong defense strategies, ensuring models remain effective even when faced with adversarial inputs.

We evaluated the HPAC-IDS and EBSNN models' robustness against PGD, FGSM, and WGAN adversarial attacks. Using an \(\epsilon\) value of 0.3, for PGD, we set an \(\alpha\) value of 0.4 and performed 20 iterations (results in Table \ref{table:res_adv}).

The HPAC-IDS exhibited significant robustness, especially for segment sizes 8 and 32, recording a 0\% severity under PGD attacks. In contrast, the segment size of 20 presented slight vulnerabilities with 5\% for PGD and 10\% for FGSM. On the other hand, the EBSNN demonstrated pronounced susceptibilities, particularly for segment sizes 8 and 20, with severity rates peaking at 95\% for PGD and 80\% for FGSM. The introduction of WGAN attacks revealed a 10\% severity for segment size 8, a 5\% severity for segment size 20, and a 15\% severity for segment size 32, signifying a modest yet discernible susceptibility. However, EBSNN showed improved resistance at segment size 32, with 5\% and 10\% severity for PGD and FGSM, but a commendable 0\% severity against WGAN for segment size 32 (as detailed in Table \ref{table:res_adv}). The cosine similarity of perturbed samples with PGD and FGSM, ranged from 0.87 to 0.96, indicating minimal deviation from the original samples. 


\newcolumntype{Y}{>{\centering\arraybackslash}X}

\begin{table}[!h]
    \centering
    \begin{tabularx}{\linewidth}{|l|Y|Y|Y|}
    \hline
    \small \textbf{Model} & \small \textbf{PGD} & \small  \textbf{FGSM}  & \small \textbf{WGAN} \\
    \hline
    \footnotesize HPAC (seg\_size 8) & \footnotesize \textbf{0\%} & \footnotesize \textbf{0\%}  & \footnotesize \textbf{10\%} \\
    \hline
    \footnotesize HPAC (seg\_size 20) & \footnotesize \textbf{5\%} & \footnotesize \textbf{10\%} & \footnotesize \textbf{5\%} \\
    \hline
    \footnotesize HPAC (seg\_size 32) & \footnotesize \textbf{0\%} & \footnotesize \textbf{0\%} & \footnotesize 15\% \\
    \hline
    \footnotesize EBSNN (seg\_size 8) & \footnotesize 95\% & \footnotesize 80\% & \footnotesize 100\% \\
    \hline
    \footnotesize EBSNN (seg\_size 20) & \footnotesize 95\% & \footnotesize  80\% & \footnotesize 65\% \\
    \hline
    \footnotesize EBSNN (seg\_size 32) & \footnotesize 5\% & \footnotesize  10\% & \footnotesize  \textbf{0\%} \\
    \hline
    \end{tabularx}
    \caption{\footnotesize Severity results under different adversarial attacks on HPAC-IDS and EBSNN models}
    \label{table:res_adv}
    \vspace{-0.6cm}
\end{table}

The HPAC-IDS model's robustness to adversarial attacks stems from its unique fusion of hierarchical attention mechanisms and convolutional neural networks (CNNs). The hierarchical attention mechanisms allocate varied attention to different network traffic segments, guarding against subtle adversarial perturbations and capturing the intricate interdependencies of network packets. Concurrently, the CNNs, known for detecting complex patterns in hierarchical data, identify subtle adversarial modifications, enhancing the model's robustness against such threats.

\section{Discussion and Conclusion} \label{section:5}


The experiments underscore the superiority of the Hierarchical Packet Attention Convolution System for Intrusion Detection Systems (HPAC-IDS) over considered existing methods. Viewing network packets akin to natural language, we introduced a \textit{Packet Segmenter} for pre-processing. The HPAC-IDS blends hierarchical attention mechanisms with convolutional neural networks, allowing it to discern complex patterns in network traffic. Considering the network communication layers' hierarchical structure equips the model with a nuanced understanding of packet interdependencies, optimizing malicious traffic detection.




The robustness of the HPAC-IDS model to PGD, FGSM, and WGAN attacks is attributed to its combination of hierarchical attention mechanisms and CNNs, which adeptly identify intricate network traffic features, ensuring robust detection against adversarial tweaks. The judicious choice of segment size further enhances the model's performance, allowing it to capture long-term traffic patterns and distinguish between normal and intrusive packets, ensuring higher detection accuracy with fewer false positives. Moreover, how we represent raw packets is pivotal; exploring approaches from time series to NLP-inspired embeddings might bolster intrusion detection system robustness against emerging threats.

\bibliographystyle{IEEEtran}
\bibliography{references}

\end{document}